\documentclass[prd,eqsecnum,notitlepage,
nofootinbib,superscriptaddress]{revtex4-1}
\global\arraycolsep=2pt
\usepackage{stmaryrd}
\usepackage{amsmath}
\usepackage{amssymb}
\usepackage{graphicx}
\usepackage{textcomp}
\usepackage{calrsfs}
\usepackage{yfonts}
\usepackage{bm}
\usepackage{color}
\newcommand{\ben}{\begin{equation*}}
\newcommand{\een}{\end{equation*}}
\newcommand{\bean}{\begin{eqnarray*}}
\newcommand{\eean}{\end{eqnarray*}}

\newcommand{\nn}{\nonumber}
\newcommand{\be}{\begin{equation}}
\newcommand{\ee}{\end{equation}}
\newcommand{\bea}{\begin{eqnarray}}
\newcommand{\eea}{\end{eqnarray}}

\DeclareMathOperator{\Tr}{Tr}

\begin{document}
\title{Remarks on the Casimir Self-Entropy of a Spherical
Electromagnetic $\delta$-Function Shell}
\author{Kimball A. Milton}
  \email{kmilton@ou.edu}
  \affiliation{H. L. Dodge Department of Physics and Astronomy,
University of Oklahoma, Norman, OK 73019 USA}
\author{Pushpa Kalauni}
  \email{pushpakalauni60@gmail.com}
  \affiliation{H. L. Dodge Department of Physics and Astronomy, University of
Oklahoma, Norman, OK 73019 USA}
\author{Prachi Parashar}
  \email{Prachi.Parashar@jalc.edu}
\affiliation{Department of Energy and Process Engineering,
Norwegian University of Science and Technology, 7491 Trondheim, Norway}
  \affiliation{John A. Logan College, Carterville, Illinois
62918, USA}
\author{Yang Li}
  \email{liyang@ou.edu}
  \affiliation{H. L. Dodge Department of Physics and Astronomy, University of
Oklahoma, Norman, OK 73019 USA}

\begin{abstract}
Recently the Casimir self-entropy of an electromagnetic $\delta$-function 
shell was considered by two different groups, with apparently discordant 
conclusions, although
both had concluded that a region of negative entropy existed for sufficiently
weak coupling.  We had found that the entropy contained an infrared divergence,
 which we argued should be discarded on physical grounds.  On the contrary, 
Bordag and Kirsten recently found a completely finite self-entropy, although 
they, in fact, have to remove an infrared divergence.  Apart from this, the
high- and low-temperature results for finite coupling agree precisely for 
the transverse electric mode, but there are significant discrepancies in the 
transverse magnetic  mode.  We resolve those 
discrepancies here.  In particular, it is shown that coupling-independent terms
do not occur  in  a consistent regulated calculation, 
they likely being an artefact of the omission of pole terms.
The results of our previous analysis, especially, the existence of a negative 
entropy region for sufficiently weak coupling,  are therefore confirmed.  
Finally, we offer some analogous remarks concerning the Casimir entropy of a 
thin electromagnetic sheet, where the total entropy is always positive.
 In that case, the origin of the analogous discrepancy can
be explicitly isolated.
\end{abstract}

\date\today
\maketitle

\section{Introduction}
\label{sec:intro}
The entropy due to electromagnetic field fluctuations, or Casimir entropy, of a
perfectly conducting spherical shell (of radius $a$) was computed 
many years ago by Balian and Duplantier \cite{bd}, who found the following low 
and high temperature behaviors for the free energy,
\be
\Delta F_\infty\sim-\frac{(\pi a)^3}{15}T^4,\quad aT\ll1; \qquad
F_\infty \sim -\frac{T}4(\ln aT+0.7686),\quad aT\gg 1.\label{hilowTsc}
\ee
Here the subscript is a reminder that the conductivity of the sphere is 
considered infinite, and the $\Delta$ means this is the correction to the 
zero-temperature Casimir energy of the sphere, first calculated by Boyer 
\cite{boyer}.  Only recently
was this calculation generalized to a spherical shell with a finite 
electromagnetic coupling, a so-called electromagnetic $\delta$-function shell, 
or a spherical plasma shell \cite{Milton:2017ghh, bordagandkirsten}.
The former is described by the background permittivity
\be
\bm{\varepsilon}(\mathbf{r})-\bm{1}=\lambda(\bm{1}-\mathbf{r r})\delta(r-a),
\ee
which describes a sphere of radius $a$ centered at the origin.  The 
anisotropy is required by Maxwell's equations, as detailed in 
Refs.~\cite{Parashar:2012it,Parashar:2017sgo}.
We further assume that the medium is dispersive, with a plasma-model like
dispersion relation, $\lambda=\lambda_0/(\zeta^2 a)$, with $\lambda_0$ a
dimensionless constant, in terms of the imaginary frequency $\zeta$.  
This model is approximately realistic, and the transverse electric (TE)
mode in this model coincides with the analogous scalar field model.  It also
coincides with the plasma-shell model considered by Bordag and Kirsten
 in Ref.~\cite{bordagandkirsten}.
To translate parameters in the model in that reference to ours, we note that
their $R$ is the same as our $a$, and $\Omega R$ coincides with $\lambda_0$.
When $\lambda_0\to\infty$ we recover the perfectly conducting spherical shell.

In this note we will make a detailed comparison between the results found in 
Refs.~\cite{bordagandkirsten} and  \cite{Milton:2017ghh}.  We will see that the
finite coupling results found at low and high temperature agree for the 
TE mode, which is by far easier to treat.  
There are some discrepancies in the transverse magnetic (TM) contributions to 
the entropy.  We see no sign of the coupling-independent high-temperature TM 
term in the free energy found in Ref.~\cite{bordagandkirsten}; this arises 
because the heat-kernel  approach incorrectly incorporates $\lambda^0$ terms,  
apparently due to the omission of a pole term in the frequency
integration.  However, the high-temperature
term linear in the coupling coincides with our findings, and results from the
exact treatment of the $O(\lambda_0)$ terms.  At low temperature, 
Ref.~\cite{bordagandkirsten} gives only the result for $\lambda_0\gg(aT)^2$, 
that is, for the temperature being the smallest scale in the problem; we show 
that their machinery yields our result for arbitrary values of 
$aT/\sqrt{\lambda_0}$. The low temperature behavior will be described in 
Sec.~\ref{sec:low}, while the high-temperature limit will be discussed in 
Sec.~\ref{sec:high}.  Finally, we note that we disagree with
their procedure of subtracting the leading high temperature terms in the free
energy; doing so would violate  the strong-coupling limit given in 
Eq.~(\ref{hilowTsc}), which we reproduce but was initially  unmentioned, 
except at zero temperature,  in Ref.~\cite{bordagandkirsten}.
Indeed, in the revised version of \cite{bordagandkirsten} they perform a 
different subtraction for the perfectly conducting sphere, so a smooth limit is
not possible.

Details of the new calculations  for the sphere are relegated to
  Appendices \ref{appa} and \ref{appb}.  In Appendix \ref{appc}
we discuss the entropy of a flat electromagnetic sheet which we considered 
earlier in Ref.~\cite{Li:2016oce},  and has now been revisited by Bordag 
\cite{bordagfs}.  Again, there is disagreement about the coupling-independent 
term, this time in the TE mode, as well as about what is to be subtracted.  
This changes the physical conclusions, in that  we find the total entropy to be
 always positive; and the total entropy for a perfectly conducting sheet is 
zero. Mathematically one can see  essential agreement of all the terms found in
 the two approaches.  In particular, in Appendix \ref{appd} we show how our 
result  is reproduced using the Abel-Plana formula, which yields a expression
very similar to that seen in Bordag's paper \cite{bordagfs}, differing
only by a crucial extra term.  The latter is the origin of the discrepant
coupling-independent term.  In Appendix \ref{appe} we identify the exact
origin of this  discrepancy: 
In the passage from the real-frequency
expression for the entropy to that obtained from
the phase-shift expression used in 
Ref.~\cite{bordagfs}, a pole contribution was omitted. 
 (Such seems to have been done in Ref.~\cite{bordagfs}, 
as we also show in Appendix \ref{appe}.)
We believe a similar omission occurs in the sphere calculation, although 
because of its greater complexity, it is less transparent to detect.

\section{Low Temperature Regime of the Free Energy}
\label{sec:low}
The leading low temperature correction given by Ref.~\cite{bordagandkirsten} is
in our notation (disregarding the subtraction of the high-temperature 
contribution, to which we return later)
\be
F_{T\to0}=-\frac{\pi^3}{15}\frac{6+\lambda_0}{3+\lambda_0}a^3T^4
=\frac{(\pi a)^3 T^4}{15}\left(\frac1{1+3/\lambda_0}-2\right),\label{bklow}
\ee
where the first term is the TE contribution and the second the TM.  The TE term
in the free energy is precisely that given in Ref.~\cite{Milton:2017ghh}, see 
Eq.~(6.3) there.  The second term is the TM free energy found there as well, 
see Eq.~(6.13), if $aT\ll\sqrt{\lambda_0}$, that is, if the dimensionless 
temperature $aT$ is the smallest quantity in the problem.  However, if this is 
not the case, there are corrections parameterized by 
$\xi=\frac{\alpha}{\sqrt{2\lambda_0/3}}$, where we've introduced
the abbreviation $\alpha=2\pi a T$.   We obtained closed form expressions for 
the TM free energy for low temperatures as a function of $\xi$, see 
Eq.~(\ref{mkpllowt}).  For small $\xi$ the result coincides with that contained
in Eq.~(\ref{bklow}),
\be
\xi\ll1:\quad F^{\rm TM}\sim -\frac{2}{15}(\pi a)^3T^4,\label{ftmlow}
\ee
which is Eq.~(6.22) of Ref.~\cite{Milton:2017ghh}, while for large $\xi$ the 
result coincides with the high-temperature limit of the exact result for the TM
 free energy in $O(\lambda_0)$:
\be
\xi\gg1:\quad F^{\rm TM}\sim \frac{2}{9}\lambda_0\pi a T^2,\label{ftmlow2}
\ee
as stated in Eq.~(6.23) of Ref.~\cite{Milton:2017ghh}.  This implies negative
entropy occurs for small coupling and temperature.  (The TE contribution is
always negative.)

These results may be easily reproduced using the methods of 
Ref.~\cite{bordagandkirsten}. The details are given in Appendix \ref{appa}.

\section{High Temperature Regime of the Free Energy}
\label{sec:high}
Here there seems more discrepancy between the two approaches, but again the 
results coincide for the TE mode.  We both have for large $aT$ (fixed 
$\lambda_0$) that [Eq.~(7.17) of Ref.~\cite{Milton:2017ghh}]
\be
aT\gg1:\quad F^{\rm TE}\sim \frac{\lambda_0\pi a T^2}6,
\ee
which results from the exact free energy in  the lowest-order in $\lambda_0$.
On the other hand, Ref.~\cite{bordagandkirsten} gives
\be
aT\gg 1:\quad F^{\rm TM}\sim -2\zeta(3)a^2 T^3+\frac{\lambda_0\pi a T^2}{18}.
\label{hiTM}
\ee
The second term  is the same as  
the high temperature limit again of the 
$O(\lambda_0)$ term given in Eq.~(7.30) of Ref.~\cite{Milton:2017ghh}, 
but we saw no evidence of the first term in Eq.~(\ref{hiTM}), which seems 
counterintuitive because it persists even if the coupling goes to zero.  But we
did not examine the general high temperature result for fixed $\lambda_0$ in 
our earlier paper, but only in the strong coupling (perfectly conducting) 
limit.  We remedy that deficiency now in Appendix \ref{appb},
and again only find the term of $O(\lambda_0)$ in Eq.~(\ref{hiTM}). 
This again implies negative entropy occurs even at high temperature for
sufficiently small coupling.  
The reason for the discrepancy with the result of Ref.~\cite{bordagandkirsten},
which was calculated by a rather elaborate method in 
Ref.~\cite{bordagandkushnutdinov}, is that we used the exact uniform asymptotic
expansion (UAE) for Euclidean frequencies together with the rapidly convergent 
Chowla-Selberg formula \cite{elizalde1,elizalde2}, 
so a term independent of $\lambda_0$ cannot occur in our calculation.

Indeed, we can  recover  a term of the same form as the first term
in Eq.~(\ref{hiTM}) by including, erroneously, a $k=0$ term in 
Eq.~(\ref{uaest}),  with the leading asymptotic term given by 
Eq.~(\ref{csest}). Evidently the approach used in Ref.~\cite{bordagandkirsten} 
does not correctly omit the $(\lambda_0)^0$ contribution from the free energy. 
This is further elucidated in the flat sheet case in
 Appendices \ref{appd} and \ref{appe}; 
in the former, we show that the Abel-Plana formula, which recasts our
Euclidean approach into real frequencies, yields our, not Bordag's, result,
and in the latter, we identify the pole term that transforms Bordag's free
energy into ours.

\section{Discussion}
\label{sec:disc}
Therefore, we have shown substantial agreement between the results of 
Refs.~\cite{bordagandkirsten} and \cite{Milton:2017ghh}, for the free
energy of a $\delta$-function sphere.  The agreement is 
perfect for the TE mode.  The TM mode is more subtle.  There, at low 
temperature, the calculations agree if the temperature (in units of the
inverse radius of the sphere) is the smallest quantity, but we point out that 
there are interesting corrections if $\lambda_0/(aT)^2$ is small, resulting in 
a sign change in the entropy. At high temperature, again we exactly agree with 
the term linear in the coupling, but we see no evidence of a term in the free 
energy, independent of $\lambda_0$, proportional to $T^3$.  We believe this
term is an artefact of the method employed by Bordag et al.  
 In the case of a flat sheet, the Abel-Plana formula, which we 
would expect to yield results equivalent to the heat-kernel approach used in 
Ref.~\cite{bordagfs}, in fact resums the free energy into a form which 
does  yield our  weak-coupling expansion \cite{Li:2016oce}.  This is  discussed
in Appendix \ref{appd}. We identify the extra pole term that resolves this
discrepancy in Appendix \ref{appe}; we expect a similar resolution in the
sphere case, but the analysis is more involved there.
Refs.~\cite{Milton:2017ghh,Li:2016oce} use temporal and
spatial point-splitting, permitting weak- and strong-coupling expansions. 
Working with Euclidean frequencies removes ambiguities in the branch lines of 
the square roots.

Ref.~\cite{bordagandkirsten} does not make any comparison of their results with
ours.  This is surprising, but they justify this by remarking that our 
procedure results in  some divergent terms.  However, at the end of the 
calculation there was only an infrared-sensitive term
\be
F^{\rm TM}_{\rm IR}= T\ln \frac\mu{T}.
\ee
This we argued should be removed as an irrelevant contact term, since it does 
not refer to the sphere parameters, and indeed precisely such a term can be 
seen to be removed implicitly in the calculation given in 
Ref.~\cite{bordagandkirsten}, as one can verify by examining the arguments
in Ref.~\cite{bordagandkushnutdinov}.

Finally, we must address the subtraction procedure advocated in 
Ref.~\cite{bordagandkirsten}. The argument given there is that the two leading 
high-temperature terms  seen in Eq.~(\ref{hiTM}) should be subtracted
because they do not possess a classical limit.  But doing so would seem to 
challenge the self-consistency of  the theory, and would result in changing the
 well-established perfectly conducting sphere limit, which is indeed 
acknowledged in the revised version of their paper \cite{bordagandkirsten}.  
Subtracting their leading, coupling-independent, term from the free energy 
further introduces an explicitly negative entropy term for weak coupling.

Both calculations discussed in this note find that there is a negative entropy 
region, which seems in contradiction with the physical, thermodynamical meaning
of entropy.  However, as Ref.~\cite{bordagandkirsten} seems to acknowledge, 
neither of us are accounting for the complete physical system.  The background,
in our case, the $\delta$-function potential, and in their case, the plasma 
shell, are established by forces  other than those arising from  the 
electromagnetic fluctuating fields  whose effects 
we are calculating.  A thorough 
investigation including the complete physical system would yield a positive
total entropy.

As we were completing  the first version of this note, Bordag 
posted a new paper \cite{bordagfs} which discusses the electromagnetic thin 
sheet, which we had considered earlier in Ref.~\cite{Li:2016oce}.  
 As we have already mentioned, in Appendices \ref{appc}, \ref{appd}
 we again show essential agreement between our two approaches, although Bordag 
again finds a spurious $\lambda_0$-independent term in now the TE component 
of the free energy,  which discrepancy is resolved in Appendix 
\ref{appe}, and advocates subtractions for which we see no necessity.

\acknowledgments
We thank Michael Bordag for illuminating conversations at the Trondheim Casimir
 Effect Workshop 2018, and for sending us preliminary
versions of his papers.  We thank Steve Fulling for insightful comments.
KAM and LY acknowledge 
the financial support of the U.S. National Science Foundation,
grant No.~1707511, PP support from the Norwegian Research Council, project 
No.~250346, and LY support of the Avenir and Carl T. Bush Foundations.

\appendix
\section{The low temperature limit of the TM free energy}
\label{appa}
\begin{figure}
\includegraphics{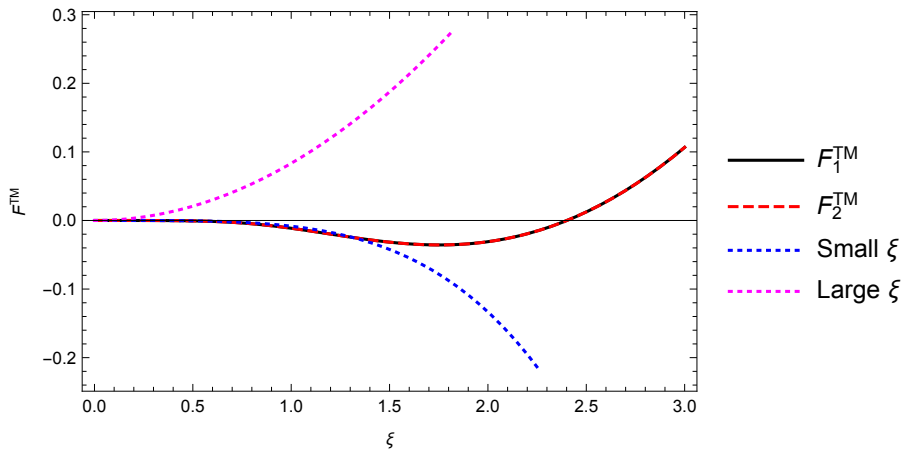}
\caption{\label{fig:lotmf} The free energy computed from either 
Eq.~(\ref{bklowt}) or Eq.~(\ref{mkpllowt}), which coincide, compared with the 
leading low $\xi$ approximation (lower dotted, blue) and with the leading large
$\xi$ approximation (upper dotted, magenta).  The latter approaches the exact 
result closely for larger $\xi$ than shown in the figure.}
\end{figure}

In this appendix we sketch how the methods of Ref.~\cite{bordagandkirsten} 
yield exactly the same result for $F^{\rm TM}$ in the low temperature limit as
found in Ref.~\cite{Milton:2017ghh}.
Bordag and Kirsten compute the free energy from the phase shifts, defined here 
by
\be
\delta_l^{\rm TM}=-\frac\pi2+\arctan\frac{1-\frac{\lambda_0}x 
\tilde{\jmath}_l'(x)
\tilde y_l'(x)}{\frac{\lambda_0}x[\tilde{\jmath}_l'(x)]^2},\quad x=\omega a,
\ee
where the Riccati-Bessel functions are defined in terms of the usual spherical 
Bessel functions by
 $\tilde{\jmath}_l(x)=x j_l(x)$, $\tilde y_l(x)=x y_l(x)$.
For small temperature, all that is relevant is the leading low frequency 
behavior, which arises only for $l=1$ (larger values of $l$ give 
higher powers of $T$):
\be
\delta_1^{\rm TM}\sim -\pi+ \frac23x^3.
\ee
From this limit the result (\ref{ftmlow}) follows.
However, if $x$ and $\lambda_0$ are comparable, there are corrections:
\be
\delta^{\rm TM}_1=-\pi+\frac23x^3\sum_{k=0}^\infty\left(\frac32
\frac{x^2}{\lambda_0}\right)^k. \label{phaseshift}
\ee
In the scheme given in Ref.~\cite{bordagandkirsten} the temperature correction 
to the free energy is given by the formula
 \be
\Delta_T F=T\int_0^\infty \frac{d\omega}{\pi}\ln\left(1-e^{-\omega/T}\right)
\frac{d}{d\omega}\delta(\omega),\quad \delta=\sum_{l=1}^\infty (2l+1)\delta_l.
\ee
 This result may be readily derived from 
the real-frequency version of Eq.~(\ref{ftmef}).
Inserting the expansion (\ref{phaseshift}) into this, we find
\be
\Delta_T F=-\frac2\pi  T\sum_{k=0}^\infty(2k+3)\left(\frac{3}{2\lambda_0}
\right)^k(aT)^{3+2k}
\Gamma(2k+3)\zeta(2k+4),
\ee
and then if we use the Euler representation of the gamma function we have
\be
\Delta_T F=-2\frac{a^3T^4}\pi\int_0^\infty dt\,t^3 e^{-t}f\left(
\frac{\xi t}{2\pi}\right), \quad \xi=\frac{2\pi aT}{\sqrt{2\lambda_0/3}},
\ee
where 
\be
f(y)=\sum_{k=0}^\infty y^{2k}\zeta(2k+4)
=\frac{3-\pi^2y^2-3\pi y\cot\pi y}{6y^4}.
\ee
This expression actually does not exist because of poles in the cotangent; 
 the radius of convergence of the series is 1.  Such poles are
characteristic of real-frequency formulations.  However, we may 
find a  unique analytic continuation 
by making,  for example, a $\pi/4$ rotation in the integration variable, 
$t\to t(1+i)$,
\be
\Delta_T F^{\rm TM}=-\left(\frac{2\lambda_0}3\right)^2\frac1{\pi a}
\frac{\xi^4}{8\pi^4}(1+i)^4\int_0^\infty dt\,t^3 e^{-t(1+i)}f\left(\frac{\xi t}
{2\pi}(1+i)\right),\label{bklowt}
\ee
which is absolutely convergent.   
This is an alternative ``closed-form'' to that shown in 
Ref.~\cite{Milton:2017ghh}, 
and gives the limits shown above in Eqs.~(\ref{ftmlow}), (\ref{ftmlow2}).
It coincides with the form given in our paper [Eq.~(6.24)] for all $\xi$
[$(\pi a)^{-1}$ was inadvertently omitted there],
\be
\Delta_T F^{\rm TM}=\left(\frac{2\lambda_0}3\right)^2\frac1{\pi a}\left[
\frac{\xi^2}{12}-\ln\xi-\Re\psi\left(1+\frac{i}\xi\right)\right],
\label{mkpllowt} 
\ee 
as seen in Fig.~\ref{fig:lotmf}, which is equivalent to Fig.~3 in 
Ref.~\cite{Milton:2017ghh}.  This further shows that the TM entropy 
(the negative
slope of the free energy with respect to $T$) is positive for strong coupling 
(small $\xi$), and negative for weak coupling (large $\xi$), with the 
transition occurring near $\xi_0=1.75\dots$.

\section{The high temperature limit of the TM free energy}
\label{appb}
The general expression for the TM component of the free energy for a 
electromagnetic $\delta$-function shell is (Eq.~(2.10) of 
Ref.~\cite{Milton:2017ghh})
\be
F^{\rm TM}=\frac{T}2\sum_{n=-\infty}^\infty \sum_{l=1}^\infty (2l+1)
\ln\left(1-\frac{\lambda_0}x e_l'(x)s_l'(x)\right),\quad x=n\alpha,\quad \alpha
=2\pi a T, \label{ftmef} \ee
where we see the appearance of the modified Riccati-Bessel functions,
\be
s_l(x)=\sqrt{\frac{\pi x}2}I_{l+1/2}(x),\quad e_l(x)=\sqrt{\frac{2x}\pi}
K_{l+1/2}(x).
\ee
To get the high-temperature behavior, it is convenient to first use the uniform
asymptotic
expansion (UAE) for the Bessel functions, 
which leads to the expansion of the logarithm appearing here:
\be
\ln\left(1-\frac{\lambda_0}x e_l'(x)s_l'(x)\right)
\sim\sum_{k=1}^\infty \frac{a^{(k)}_{\rm TM}(z)}{(2\nu)^k},\quad x=\nu z,\quad 
\nu=l+1/2,\quad t=(1+z^2)^{-1/2}.
\ee
The first four of these coefficients are given in Eq.~(7.26) of 
Ref.~\cite{Milton:2017ghh}.
The leading term is that with the highest power of $\lambda_0$ in each 
coefficient, which amounts to retaining only the leading order uniform 
asymptotic  expansion of the Bessel functions  within the logarithm.  Therefore
we approximate the increasingly elaborate structure of the expansion 
coefficients by
\be
a_{\rm TM}^{(k)}\sim (-1)^{k+1}\frac1k\left(\frac{\lambda_0}{z^2t}\right)^k.
\ee
This then leads to the approximate form (the prime designates 1/2 weight for $n
=0$)
\be
F^{\rm TM}\sim T\sum_{k=1}^\infty (-1)^{k+1}\frac1k
\sum_{n=0}^\infty{}' \sum_{l=1}^\infty 2\nu \left(\frac{\lambda_0}{2\nu z^2}
\right)^k (1+z^2)^{k/2}, \quad z=n\alpha/\nu
\ee
This is actually incorrect for $n=0$, where we should use the small argument 
expansion of the Bessel functions, as explained in Ref.~\cite{Milton:2017ghh}. 
The $n=0$ term requires an infrared regulator, and is a bit subtle, but however
it is precisely defined, it can only contribute an $O(T)$ contribution, smaller
than the leading terms we are seeking.

So to extract the leading high-temperature contributions to the free energy, we
 consider
\be
F^{\rm TM}_{n>0}\sim T\sum_{k=1}^\infty\sum_{n=0}^\infty\sum_{l=0}^\infty 
(2l+3) \frac{(-1)^{k+1}}k\left(
\frac{\lambda_0}2\right)^k\frac{[(l+3/2)^2+\alpha^2(n+1)^2]^{k/2}}{[\alpha^2
(n+1)^2]^k}.\label{leadinghiT}
\ee
Here we have shifted the $n$ and $l$ variables down by 1 to put the sum in 
standard form.  This expression does not actually exist; we will define it by 
analytically continuing the exponent in the numerator to $s-1<-1/2$, and then 
at the end continuing back to $s=1+k/2$.  We can take care of the factor of 
$2l+3$ by differentiating with respect to $b$, a variable to be set to 3/2
at the end.  In this way our approximation reads
\be
F^{\rm TM}_{n>0}\sim T\sum_{k=1}^\infty \frac{(-1)^{k+1}}k\left(
\frac{\lambda_0}2\right)^k\frac1s\frac\partial{\partial b}Z, \label{uaest}
\ee
where
\be
Z=\sum_{n,l=0}^\infty \frac{[(l+b)^2+\alpha^2d(n+1)^2]^s}{[(n+1)^2\alpha^2]^k}.
\ee
We have introduced yet another parameter $d$, to be set to 1 at the end, so 
that by differentiating with respect to $d$ we can get rid of the denominator:
\be
\left(\frac\partial{\partial d}\right)^k Z=\frac{\Gamma(s+1)}{\Gamma(s+1-k)}
E_2(k-s;1,d\alpha^2;b,1).\label{derZ}
\ee
In this expression  we have followed the notation of Elizalde 
\cite{elizalde1,elizalde2}.  
The high temperature behavior is captured by the generalized Chowla-Selberg 
formula given there [see Eq.~(7.3) of Ref.~\cite{Milton:2017ghh}):
\be
E_2(k-s;1,d\alpha^2;b,1)\sim (d\alpha^2)^{s-k}\zeta(0,b)\zeta(2k-2s),
\ee
with higher terms being down by powers of  $\alpha^{-2}$.
Then we can integrate up the derivatives seen in Eq.~(\ref{derZ}), but there 
are $k$ integration constants:
\be
Z\sim d^s\alpha^{2s-2k}\zeta(0,b)\zeta(2k-2s)+\sum_{j=0}^{k-1}A_j d^j, 
\label{csest}
\ee
where we can now set $d=1$ and $s=k/2+1$. The integration constants $A_j$ can 
be readily computed by evaluating $Z$ and its derivatives at $d=0$. 
However, these constants are innocuous for extracting the leading 
behavior:  For a given power of $\lambda_0$ in the free energy, the largest 
term in $\alpha$ comes from the $A_{k-1}$ term, which goes like $\alpha^{-2}$, 
subdominant compared to the leading terms. Therefore we disregard those terms, 
do the $b$ derivative by noticing that
\be
\frac\partial{\partial b}\zeta(0,b)\bigg|_{b=3/2}=-1,
\ee
and write  ($aT\to\infty$)
\be
F^{\rm TM}\sim-T\alpha^2\left[\frac{\lambda_0}{2\alpha}\frac23\zeta(-1)-\frac12
\left(\frac{\lambda_0}{2\alpha}\right)^2\frac12\zeta(0)
+\frac13\left(\frac{\lambda_0}{2\alpha}\right)^3\frac25\zeta(1)+
\sum_{k=4}^\infty \frac{(-1)^{k+1}}{k}\left(\frac{\lambda_0}{2\alpha}\right)^k
\frac1{1+k/2}\zeta(k-2)
\right].
\ee
The sum turns out to be of order $(\lambda_0/2\alpha)^3$; but that and the 
third divergent term are temperature independent. The second term here is of 
order $T$, but that must be supplemented  by the $n=0$ term which we deferred 
above.  Thus, all we can extract is the leading term for high temperature,
\be
F^{\rm TM}\sim \frac{\lambda_0 \pi a T^2}{18},\quad aT\gg1,\lambda_0,
\ee
which coincides with the second term in Eq.~(\ref{hiTM}), and which, as 
anticipated, agrees with the high temperature limit of the exact $O(\lambda_0)$
solution.  No sign appears of  the first term in Eq.~(\ref{hiTM}), 
which is coupling-independent.  The reason for the discrepancy
 with that of  the procedure used in Ref.~\cite{bordagandkushnutdinov} is that 
our  regulated expressions for the free energy vanish in the 
absence of interactions, so there can be no contribution at $\lambda_0=0$.  
It appears, as demonstrated in Appendix \ref{appe} for the
analogous flat sheet problem, that in translating the free energy expression
into the phase-shift formulation used in Refs.~\cite{bordagandkirsten} and
\cite{bordagfs} a pole term has been omitted, whose inclusion would cancel
the offending term.  The unregulated 
heat-kernel technique, unlike the Abel-Plana formula,  discussed in Appendix 
\ref{appd}, inserts spurious coupling-independent terms.
Further evidence for the appropriateness of our approach lies in the 
strong-coupling (perfect conductor) limit, where there is a term of 
just such a form in both the TE and TM modes, occurring
with equal magnitudes and opposite signs, so they cancel in the total free 
energy.  (This was seen also in Refs.~\cite{bkmm,geyer}.)  Here, the term
appears only in the TM mode.  Finally we note that in our procedure, detailed 
in Ref.~\cite{Milton:2017ghh}, where no such term appears, we do recover the 
Balian and Duplantier] result (\ref{hilowTsc}) for the perfect-conductor, 
high-temperature limit.  No smooth limit is possible in the scheme advocated in
 Ref.~\cite{bordagandkirsten}.

\section{Thin electromagnetic sheet}
\label{appc}
In Ref.~\cite{Li:2016oce} 
we exactly solved for the Casimir entropy of a flat
electromagnetic $\delta$-function sheet, described by the permittivity 
$\bm{\varepsilon}(\mathbf{r})-\bm{1}=\mbox{diag}(\lambda,\lambda,0)\delta(z)$. 
We showed that the TE entropy is always negative, while the TM entropy is 
positive, and always larger than the magnitude of the former. The total entropy
tends to zero in the limit $\lambda\to \infty$, that is, for a perfectly 
conducting sheet. Results were precisely defined using temporal and spatial
point-splitting regulators.

Closed form results were obtained for the entropy for a ``plasma model,'' where
 the dispersion was incorporated by writing $\lambda=\lambda_0/\zeta^2$, where 
$\lambda_0$ is a constant, and $\zeta$ is the imaginary frequency.  (For the 
flat sheet, $\lambda_0$ has the dimension of (length)$^{-1}$. 
The explicit forms for 
the TE and TM entropies per unit area are given by (4.13) and (4.25) of 
Ref.~\cite{Li:2016oce}.  We will content ourselves here by writing the limits:
\begin{subequations}
\bea
T\gg\lambda_0:\quad S^{\rm TE}&\sim& -\frac{\lambda_0}{12}T+O(T^0),
\label{TEhigh}\\
T\ll \lambda_0:\quad S^{\rm TE}&\sim& -\frac{3\zeta(3)}{4\pi} T^2+O(T^3),
\label{TElow}
\eea
\end{subequations}
and
\begin{subequations}
\bea
T\gg\lambda_0:\quad S^{\rm TM}&\sim& \frac{120\zeta(5)}{\pi\lambda_0^2}T^4+
\frac{3\zeta(3)}{2\pi}T^2 -\frac{\lambda_0}{36}T+O(T^0),\label{TMhigh}\\
T\ll \lambda_0:\quad S^{\rm TM}&\sim& \frac{3\zeta(3)}{4\pi} T^2+O(T^3).
\label{TMlow}
\eea
\end{subequations}
Notice that these results mean that the total entropy vanishes in the perfect 
conducting limit.

Ref.~\cite{bordagfs} seems to obtain somewhat different limits.  For high 
temperature, Bordag gives (with his $\Omega_0=\lambda_0/2$ and $\omega_0=0$) 
for the TE contributions,
\be
T\gg \lambda_0:\quad 
S^{\rm TE}_B\sim \frac{3\zeta(3)}{4\pi}T^2-\frac{\lambda_0}{12}T,\label{hiTTE}
\ee
so although the second term coincides with Eq.~(\ref{TEhigh}), the first term 
was not seen by us. (The corresponding heat kernel coefficients were first 
worked out in Ref.~\cite{Bordag:2005qv}.)
 Again, this is presumably because our regulated 
expressions allow for a weak-coupling expansion.  Indeed, were we to start the 
sum in Eq.~(4.11) in Ref.~\cite{Li:2016oce} at $n=-1$ ($n=0$ is already 
explicitly included), we would obtain (taking the limit $n\to-1$) 
exactly the first term in Eq.~(\ref{hiTTE}).  Again, this is clearly incorrect.
Once more, because he subtracted both of these leading terms from the entropy, 
his subtracted TE entropy  per unit area has a linear term at low temperature:
\be
T\ll \lambda_0:\quad
S^{\rm TE}_{B,\rm{sub}}\sim\frac{\lambda_0}{12} T,\label{TEsub}
\ee the term shown in Eq.~(\ref{TElow}) being of higher order.

For TM Bordag recognizes the first two terms in the high-temperature limit 
(\ref{TMhigh}) as the TM surface plasmon contributions ($\omega_0=0$), 
which he again subtracts, leaving precisely the third term there:
\be
T\gg\lambda_0:\quad S^{\rm TM}_{B,{\rm sub1}}\sim-\frac{\lambda_0}{36}T,
\ee
but he subtracts this term away as well, leaving a low-temperature entropy 
per unit area exactly one-third of that for TE in Eq.~(\ref{TEsub}):
\be
T\ll\lambda_0:\quad S^{\rm TM}_{B,\rm{sub2}}=\frac{\lambda_0}{36}T,
\ee
because again the correction from Eq.~(\ref{TMlow}) is higher order.  
Note, that with Bordag's prescription, the perfect conductor limit does not 
exist.

So the technical results of both paper coincide, as further shown in Appendix 
\ref{appd}.  We disagree only the inclusion of spurious coupling-independent 
terms, and on the necessity of subtracting terms because they do not seem to 
reproduce known results. The following two Appendices help resolve 
the issue of  the spurious terms.

\section{Abel-Plana formula}
\label{appd}
For simplicity, we consider here the TE mode of the free energy per area for 
the thin sheet, which is given by the  spatially regulated formula
(Eq.~(4.1) of Ref.~\cite{Li:2016oce})
\be
F^{\rm TE}=\frac{T}{2\pi}\sum_{m=0}^\infty{}'\int_0^\infty dk \,k
J_0(k\delta)\ln\left(1+\frac{\lambda_0}
{2\kappa_m}\right),\quad \kappa_m=\sqrt{k^2+\zeta_m^2},\label{fesheet}
\ee
where the prime means the $m=0$ term is counted with half weight.
The Abel-Plana formula allows us to turn the sum into an integral:
\be
\sum_{m=0}^\infty{}'f(m)=\int_0^\infty dt\,f(t)+i\int_0^\infty dt
\frac1{e^{2\pi t}-1}[f(it)-f(-it)].
\ee
Using the first term here in the formula for the free energy (\ref{fesheet}) 
gives a term independent of temperature, which we disregard.  For the second 
term we integrate by parts
\be
\Delta F^{\rm TE}=\frac{T}{(2\pi)^2}\int_0^\infty dt\ln(1-e^{-2\pi t})[f'(it)
+f'(-it)],
\ee
where
\be
f(t)=\int_0^\infty dk\,k J_0(k\delta)
\ln\left(1+\frac{\lambda_0}{2\sqrt{k^2+(2\pi t T)^2}}\right).
\ee
 The derivative of $f$ does not require the regulator:
\be
f'(t)=(2\pi T)^2t\ln\left(1+\frac{\lambda_0}{4\pi t T}\right).
\ee
and then
with  $\tilde\omega=2\pi t$, we have
\be
f'(it)+f'(-it)
=-4\pi  T^2\tilde\omega\arctan\frac{\lambda_0}{2\tilde\omega T}.
\ee
In this way we obtain  a result slightly  different from what Bordag gives:
\be 
\Delta F^{\rm TE}=-\frac{T^3}{2\pi^2}\int_0^\infty d\tilde\omega\,\tilde
\omega\ln\left(1-e^{-\tilde\omega}\right)\arctan\frac{\lambda_0}{2\tilde
\omega T},
\label{bordagsfesheet}
\ee
while Ref.~\cite{bordagfs} has the same formula with the arctangent replaced
by $-\arctan\frac{2\tilde\omega T}{\lambda_0}=\arctan\frac{\lambda_0}{2
\tilde\omega T}
-\frac\pi2$.

If we expand the arctangent for large argument,
 we obtain the nearly the same leading high-temperature
result that Bordag does:
\be
\frac{T}{\lambda_0}\gg1:\quad
\Delta F\sim
\frac{\lambda_0}{24}T^2-\frac{\lambda_0^2}{24}T
\left(1-2\ln\frac{\lambda_0}{2T}\right),
\ee
which is consistent with Eq.~(\ref{hiTTE}), 
apart from the first term there.
The  two  terms here agree with those found in 
Eq.~(4.14b) of Ref.~\cite{Li:2016oce},  and, as shown there, the full
series is convergent.  In the opposite limit, that of low temperature or
strong coupling, we obtain from Eq.~(\ref{bordagsfesheet}) the asymptotic 
series
\be
\frac{\lambda_0}{T}\gg1:\quad
\Delta F^{\rm TE}= -\frac{\zeta(3)}{4\pi}T^3
-\frac{T^3}{2\pi^2}\sum_{n=0}^\infty (-1)^n\frac{\Gamma(2n+4)\zeta(2n+4)}
{(2n+1)(2n+3)}\left(\frac{2T}{\lambda_0}\right)^{2n+1},
\ee 
which coincides with our expansion (4.14a) of Ref.~\cite{Li:2016oce}.
In general, we conclude that the difference between the two forms of the 
entropy is
\be
S^{\rm TE}=S^{\rm TE}_{\rm B}-\frac{3\zeta(3)}{4\pi}T^2.
\ee
This suggests that
that the properly regulated theory is that discussed in Ref.~\cite{Li:2016oce},
so the coupling-independent term is not present. 
 This is demonstrated in the following Appendix.

\section{Resolution of Discrepancy}
\label{appe}

We now carefully rederive the starting point in Ref.~\cite{bordagfs} starting
from the  real-frequency form for the free energy ($\beta=1/T$),
which follows directly from the familiar $\Tr\ln$ formula
$F=-\frac12\Tr\ln\bm{\Gamma}\bm{\Gamma}_0^{-1}$ in terms of the free and
full Green's dyadics $\bm{\Gamma}_0$ and $\bm{\Gamma}$.  For the
transverse electric contribution to the free energy this amounts to
\bea
\Delta F^{\rm TE} &=&\frac1{2\pi^2}\Im\int_0^\infty d\omega
\frac1{e^{\beta\omega}-1}\int_0^\infty
dk\,k\ln\left(1+\frac{\lambda_0}{2\sqrt{k^2-\omega^2}}\right)\nn\\
&=&-\frac{\lambda_0}{4\pi^2\beta}\Im \int_0^\infty dk\,k\int_0^\infty d\omega
\,\omega \ln\left(1-e^{-\beta\omega}\right)\frac1{k^2-\omega^2}
\frac1{\lambda_0/2+\sqrt{k^2-\omega^2}}.\label{realfe}
\eea
(Formally, this can be derived from the Euclidean form
(\ref{fesheet}) by the Abel-Plana formula.)
In the second line, we have integrated by parts and omitted the boundary
term because it is real.  There are two singularities in the $\omega$ 
integration above, a pole and a branch point, both occurring at $\omega=k$.
We choose the branch line to pass from $k$ to $\infty$.
In the spirit of using the causal or Feynman propagator, our contour of 
integration must pass above all of these singularities.  Let us change
variables from $\omega$ to $\kappa=\sqrt{k^2-\omega^2}$, where $\kappa$ is
real for $\omega<k$ and $\kappa=-i k_z$ for $\omega>k$, the sign of $i$ being
dictated by the above contour requirement.  We then write the free energy as
\be
\Delta F=-\frac{\lambda_0}{4\pi^2\beta}\Im\int_0^\infty
dk\,k\left\{\int_0^k
\frac{d\kappa}\kappa\frac{\ln\left(1-e^{-\beta\sqrt{k^2-\kappa^2}}\right)}{
\lambda_0/2+\kappa}-\int_0^\infty\frac{dk_z}{k_z}\frac{\ln\left(1-
e^{-\beta\sqrt{k^2+k_z^2}}\right)}{\lambda_0/2-ik_z}\right\}.\label{fullfe}
\ee
We will initially disregard the pole at $\kappa=k_z=0.$  Then the first term
in the above is purely real, so it is to be discarded, and the imaginary
part of what is left is
\be
\Delta_1 F=\frac{\lambda_0 T}{4\pi^2}\int_0^\infty dk\,k\int_0^\infty dk_z
\frac1{k_z^2+(\lambda_0/2)^2}\ln\left(1-e^{-\beta\sqrt{k^2+k_z^2}}\right).
\label{bordagsres1}
\ee
This is precisely the formula (10) given in Ref.~\cite{bordagfs},
with the derivative of the phase shift (or the density of states factor)
given there by 
\be
\frac{d}{dp}\delta(p)=\frac{\Omega_0}{\Omega_0^2+p^2},\label{derps}
\ee
which coincides with Eq.~(30) of Bordag's paper (with the first $\omega_0$
in the denominator replaced by $\Omega_0$.) (Remember our translation of
variables: here $\Omega_0\to\lambda_0/2$, $p\to k_z$, and $\omega_0=0$.)
This then leads directly, upon introducing polar coordinates,  
to Bordag's results for the free energy and entropy:
\be
\Delta_1 F=\frac{T}{2\pi^2}\int_0^\infty d\omega\,\omega\ln\left(1-e^{-\beta
\omega}\right)\arctan\frac{2\omega}{\lambda_0}.\label{bordagsres2}
\ee
Let us now include the pole terms that we omitted following
Eq.~(\ref{fullfe}). This gives another contribution to the imaginary part:
\be
\Delta_2 F=-\frac{\frac\pi 2}{2\pi^2\beta}\int_0^\infty dk\,k\ln
\left(1-e^{-\beta k}\right).\label{delta2f}
\ee
Combining this with the $\Delta_1 F$ contribution yields our result 
(\ref{bordagsfesheet}).

In fact, Bordag's starting point \cite{bordagfs}
\be
\Delta F^{\rm TE}=\int_0^\infty \frac{dp}\pi \int \frac{(d\mathbf{k})}
{(2\pi)^2} T\ln\left(1-e^{-\beta\omega}\right)\frac{d\delta(p)}{dp},
\ee
properly interpreted, also yields the same result.  This is because
\be
\frac{d\delta(p)}{dp}=\Re \frac{\Omega_0}{p(p-i\Omega_0)}
\ee
is not exactly Eq.~(\ref{derps}) because $p$ contains an implicit branch line,
with branch point at $p=0$.  Thus
\be
\Delta F^{\rm TE}=-T\Im\int_0^\infty \frac{dp}\pi\int \frac{dk\,k}{2\pi}
\ln\left(1-e^{-\beta\omega}\right)\left(\frac1p-\frac1{p-i\Omega_0}\right).
\ee
The second term is Bordag's result (\ref{bordagsres1}) and (\ref{bordagsres2}),
while the first, evaluated by integrating over a quarter circle around
the pole at $p=0$ in the positive sense, yields precisely our correction
(\ref{delta2f}).  (The sense of the contour is most easily seen starting from
Eq.~(\ref{realfe}).)

The discrepancy is thus resolved.  We presume a similar extra term occurs
in the more complicated spherical calculation.

\end{document}